\documentclass[USenglish,oneside,twocolumn]{article}

\usepackage[utf8]{inputenc}
\usepackage[big]{dgruyter_NEW}

\graphicspath{{figures/}}

\usepackage{hyperref}
  \hypersetup{pdfpagemode=UseNone}
\usepackage{lipsum}
\usepackage{color}
  \definecolor{blue}{RGB}{0,0,250}
  \definecolor{red}{RGB}{250,0,0}
\usepackage{comment}
\usepackage{array}
  \newcolumntype{x}[1]{>{\centering\let\newline\\\arraybackslash\hspace{0pt}}p{#1}} 
\renewcommand{\arraystretch}{1.2} 

\DOI{foobar}
  
\begin{document}

\author*[1]{Vincent F. Taylor}
\author[2]{Alastair Beresford}
\author[3]{Ivan Martinovic}

\affil[1]{Department of Computer Science, University of Oxford, Oxford, United Kingdom. E-mail: vincent.taylor@cs.ox.ac.uk}
\affil[2]{University of Cambridge, Cambridge, United Kingdom. E-mail: alastair.beresford@cl.cam.ac.uk}
\affil[3]{Department of Computer Science, University of Oxford, Oxford, United Kingdom. E-mail: ivan.martinovic@cs.ox.ac.uk}

\title{\huge Intra-Library Collusion: A Potential Privacy Nightmare on Smartphones}
\runningtitle{Intra-Library Collusion: A Potential Privacy Nightmare on Smartphones}


\begin{abstract}
{Smartphones contain a trove of sensitive personal data including our location, who we talk to, our habits, and our interests. Smartphone users trade access to this data by permitting apps to use it, and in return obtain functionality provided by the apps. In many cases, however, users fail to appreciate the scale or sensitivity of the data that they share with third-parties when they use apps. To this end, prior work has looked at the threat to privacy posed by apps and the third-party libraries that they embed. Prior work, however, fails to paint a realistic picture of the full threat to smartphone users, as it has typically examined apps and third-party libraries in isolation.\\
\mbox{\hspace{0.5cm}}In this paper, we describe a novel and potentially devastating privilege escalation attack that can be performed by third-party libraries. This attack, which we call \textit{intra-library collusion}, occurs when a single library embedded in more than one app on a device leverages the combined set of permissions available to it to pilfer sensitive user data. The possibility for intra-library collusion exists because libraries obtain the same privileges as their host app and popular libraries will likely be used by more than one app on a device.\\
\mbox{\hspace{0.5cm}}Using a real-world dataset of over 30,000 smartphones, we find that many popular third-party libraries have the potential to aggregate significant sensitive data from devices by using intra-library collusion. We demonstrate that several popular libraries already collect enough data to facilitate this attack. Using historical data, we show that risks from intra-library collusion have increased significantly over the last two-and-a-half years. We conclude with recommendations for mitigating the aforementioned problems.}
\end{abstract}


\DOI{}

\startpage{1}


\maketitle

\section{Introduction}

The average smartphone has more than 25 apps installed~\cite{nielsonmanyapps}, each having varying access to the device depending on the permissions that have been granted to them. These permissions are shared by third-party libraries (henceforth called \textit{libraries}, for brevity), which are embedded within apps and enjoy the privileges granted to the host apps. Libraries are used extensively across the app ecosystem. App developers use them to monetize apps, integrate with social media, or simply to provide complex app functionality with little programming effort required. While libraries provide unquestionable benefit to app developers and the app ecosystem as a whole, prior work has shown that they also contribute negative effects to user privacy~\cite{DBLP:journals/corr/abs-1303-0857, Grace:2012:UEA:2185448.2185464, stevens2012investigating, seo2016flexdroid}. For example, libraries may track users, abuse the permissions they have been granted, or leak sensitive personal data.

The Android security model does not support the separation of privileges between apps and their embedded libraries. As such, not only do libraries inherit the permissions granted to their host apps, the developers of the host apps themselves are sometimes forced to declare additional permissions to support embedded libraries~\cite{Shekhar:2012:ASS:2362793.2362821, Pearce:2012:APS:2414456.2414498}. Additional permissions may especially benefit advertising libraries (henceforth called \textit{ad libraries}) as they facilitate behavioural advertising through greater potential for user profiling. Ur et al.~\cite{Ur:2012:SUS:2335356.2335362} found that users were generally unaware of the inner workings of behavioural advertising, and described the practice as ``scary'' and ``creepy''. Along similar lines, Spensky et al.~\cite{spensky2016sok} argue that users cannot be expected to fully understand the implications of sharing their personal data.\\

\begin{figure*}[!t]
\centering
\includegraphics[width=6.7in,clip=true,trim=0.2in 8.2in 1.6in 0.4in]{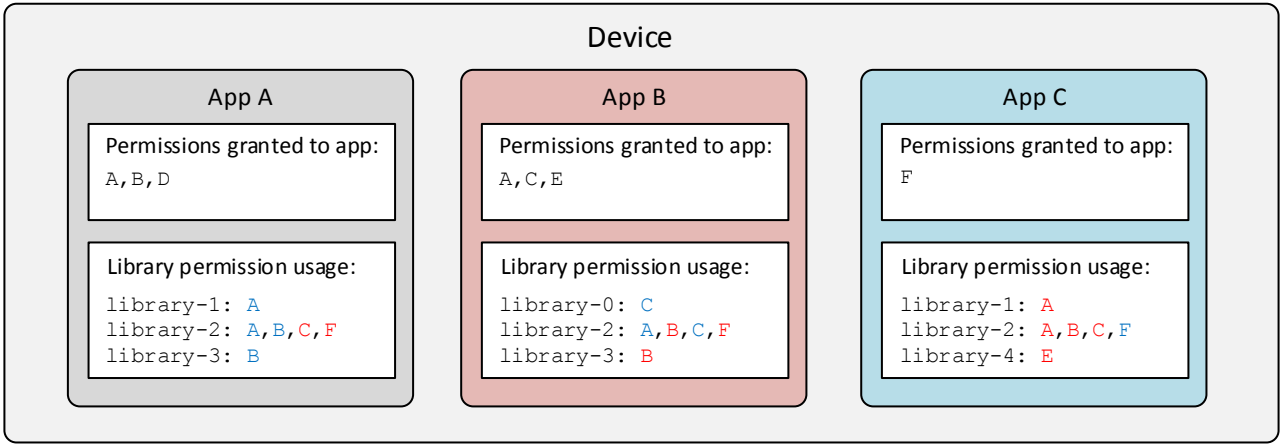}
\caption{An example of how intra-library collusion (ILC) could happen in practice. Libraries are able to use permissions in \textcolor{blue}{blue} because they have been granted to the app, while permissions in \textcolor{red}{red} are unavailable for libraries within that app. Overall, \texttt{library-2} is able to access a total of four permissions on the device.}
\label{fig-ilc-explained}
\end{figure*}

The current Android security model facilitates the following threats from libraries:\\

\begin{itemize}
\item Libraries may abuse the privileges granted to host apps.
\item Libraries may track users without their consent.
\item Opportunistic libraries may aggregate multiple signals for detailed user profiling.
\end{itemize}

Recently, Bosu et al.~\cite{Bosu:2017:CDL:3052973.3053004} studied the possibility of more than one apps colluding to leak sensitive data using inter-component communication (ICC). Prior studies have also examined privacy leaks using ICC~\cite{Lu:2012:CSV:2382196.2382223, Felt:2011:PRA:2028067.2028089, ndsssdcl, Chan:2012:DAA:2185448.2185466}, but real-world attacks are limited since they require one\footnote{One malicious app is sufficient for a confused deputy attack~\cite{bugiel2012towards} if a benign, but vulnerable, app is present on the device.} or more malicious apps to be installed on devices to facilitate the attack.

On the other hand, a novel privilege escalation attack we call \textit{intra-library collusion} (ILC) is much more likely. ILC is the phenomenon that happens when individual libraries obtain greater combined privileges on a device by virtue of being embedded within multiple apps, with each app having a distinct set of permissions granted. The ``collusion'' between instances of the same library need not happen on the device. Indeed, the library can transmit individual streams of sensitive user data to a remote server where it can be aggregated to better profile the unaware user. Prior work~\cite{stevens2012investigating} has only looked at the possibility of tracking users across apps, whereas ILC concerns aggregating sensitive user data across apps.

ILC is visually depicted in Fig.~\ref{fig-ilc-explained}. From the figure, \texttt{library-2} is shared across all apps, and has combined access to four permissions (\texttt{A,B,C,F}), although in any single app it has access to a maximum of two permissions (\texttt{A,B} or \texttt{A,C}). Thus, \mbox{\texttt{library-2}} may benefit from ILC. Note that libraries may also further their malfeasance by leveraging sources of data not guarded by permissions, such as the list of apps installed on a device. Such data has previously been shown to be useful for predicting user traits~\cite{Seneviratne:2014:PUT:2636242.2636244}.

Concretely, ILC is possible if the following conditions are true:\\

\begin{enumerate}
\item Two or more installed apps, $A_1, ..., A_n$, contain the same library $L$. 
\item Apps $A_1, ..., A_n$ have sets of permissions granted to them such that the union of the sets of permissions $U$ contains more permissions than any of the individual sets of permissions. 
\item $L$ contains code that allows it to make API calls that are guarded by two or more permissions in $U$.
\item $L$ has the ability to uniquely identify a device.
\item $L$ transmits sensitive data to its server.
\end{enumerate}

Privilege escalation using ILC is a greater problem than privilege escalation through ICC, since libraries are designed to be embedded within multiple apps by definition. This increases the likelihood that any given device will contain two or more apps with the same library embedded. Moreover, the existence of very popular libraries with total install-bases in the billions suggests that potential for ILC should be very common in practice. To make matters worse, libraries may have legitimate claims as to why personal data needs to be leaked from a device, but any aggregation of this data once it gets to library servers will be opaque to users and industry regulators alike.

Risks coming from traditional libraries are not the only worry. Chen et al.~\cite{7546512} argue that libraries are already being repackaged for propagating malicious code. Using their methodology, the authors found that 6.84\% of apps obtained from the Google Play Store were infected with potentially harmful libraries. Potentially harmful behaviour was also seen in the iOS versions of libraries. These observations demonstrate that attackers have turned their attention to libraries as a means of malware propagation, further motivating the need to study ILC and its effects.

Measuring the potential for ILC requires knowledge of the lists of apps installed on real-world devices. To obtain lists of installed apps on real-world devices, we interfaced with the Device Analyzer project~\cite{Wagner:2015:DAP:2766498.2774992}. The Device Analyzer project is concerned with obtaining usage data and other metadata from smartphones, including lists of installed apps on devices. From the Device Analyzer dataset, we were able to obtain the lists of apps installed on over 30,000 real-world devices to facilitate our study. In contrast to prior work, we measure real-world privacy risks by leveraging the additional insight obtained from access to full lists of apps that are installed on devices.

In this work, we focus on Android due to the availability of data on lists of apps installed on Android devices. However, due to similarities in access control and app deployment on iOS, we believe many of our insights may hold on that platform as well.\\

\noindent{\textbf{Contributions.} Our paper makes the following contributions to the state-of-the-art:\\}

\begin{itemize}
	\item We describe a novel privilege escalation attack called intra-library collusion (ILC), which comes from libraries embedded within apps.
	\item We perform the first study on the extent to which libraries may abuse ILC for better user profiling in the real-world~(Section~\ref{section-ilc}).
	\item We use a historical dataset of apps to demonstrate that the potential for ILC (and its consequences) have increased over the last two-and-a-half years~(Section~\ref{section-longitudinal}).
	\item We make a first effort to systematically measure the frequency of transmission of personal data from real-world devices by advertisement libraries~(Section~\ref{section-frequency-leaks}).
	\item To justify one of our experimental assumptions, we measure the adoption of run-time permissions across popular apps in the Google Play Store~(Section~\ref{section-runtime-permissions}).
\end{itemize}

\noindent{\textbf{Roadmap.} The rest of the paper is organised as follows: Section~2 gives background on important Android-related concepts; Section~3 describes the methodology that was used for data collection and analysis; Section~4 presents our measurements of the potential for ILC; Section~5 shows how the potential for ILC has increased over time; Section~6 studies how often and to how many different destinations sensitive data is leaked; Section~7 discusses the results of our study; Section~8 overviews related work, including mitigation strategies; and finally Section~9 concludes the paper.}
\section{Background}

We frame the problem of ILC by first describing the Android permission model, explaining how libraries fit into the ecosystem, and explaining our threat model.

\subsection{Android Permissions Model}

The Android operating system (OS) mediates access to sensitive device resources using a permission-based access control system. Permissions are divided into \textit{normal} and \textit{dangerous} permissions, with normal permissions protecting resources that have very little risk to user privacy and dangerous permissions guarding access to private user information and data~\cite{androiddangerousperms}. Apps wanting access to sensitive resources guarded by dangerous permissions must request the relevant permission and have it granted by the user. Throughout this paper, we focus exclusively on dangerous permissions. For this reason, we hereafter refer to \textit{dangerous permissions} as \textit{permissions}. 

Apps list the permissions they would like to use in a manifest file that is bundled within the app. Prior to Android~6.0, the permissions listed by an app had to be accepted in full at install-time in an all-or-nothing manner. As of Android~6.0, permissions must still be listed in their entirety in the app's manifest, but they may be accepted or rejected selectively by users at run-time. In order for run-time permissions to be triggered, however, the app itself must have a \texttt{targetSDK} of~23 or higher.

Android apps are packaged as \texttt{apk} files, which are compressed archives containing all the resources required by the app. This means that any libraries leveraged by an app are compiled into the same app binary for distribution. The Android OS does not provide a facility for privilege separation between apps and their embedded libraries and thus libraries inherit all permissions granted to apps. Conversely, many libraries use permissions that are not required by the app that embeds them, but due to the lack of privilege separation, apps also gain additional privileges. This is one explanation as to why many functionally-similar apps ask for greatly differing permissions~\cite{Taylor:2016:SSP:2994459.2994474}. The larger problem of privilege separation between apps and their libraries have been explored by several authors~\cite{Pearce:2012:APS:2414456.2414498, Shekhar:2012:ASS:2362793.2362821, seo2016flexdroid}.

\subsection{Embedded Libraries}

The problem of the lack of privilege separation in apps benefits libraries in general and ad libraries in particular. Ad libraries are provided by ad networks, which serve as the middle-man connecting advertisers to app developers. Ad libraries provide easy-to-use interfaces that allow app developers to quickly insert ads into their existing apps. The ad library does the heavy lifting by fetching, displaying, and tracking revenue from ads.

We suspect that ad libraries stand to benefit more from ILC than other libraries because additional ``signals'' obtained from greater access to permissions directly benefit their ability to perform behavioural and demographic targeting of ads. This has an end result of them being better able to provide targeted ads, which increases click-through rates and consequently generates increased profits. This is one explanation for the observations in prior work~\cite{Grace:2012:UEA:2185448.2185464, stevens2012investigating}, where ad libraries check for and use undocumented permissions, i.e., permissions that have not been described in their SDK documentation. Other libraries may also benefit from added access to permissions, as any data they collect can be aggregated and sold to third-parties.

Since libraries may be installed in a variety of apps with varying permissions, they typically contain code that checks whether the relevant permissions have been properly granted before making a permission-protected API call~\cite{Grace:2012:UEA:2185448.2185464, stevens2012investigating}. Indeed, Fang et al.~\cite{Fang:2016:RCA:2897845.2897914} observed that libraries were better designed than apps on a whole as it relates to handling cases where permissions have been revoked or not granted. 

In ad libraries, the use of any permissions should be carefully scrutinised, since the sole purpose of the ad library is to serve ads. That is, the ad library itself provides no other functionality on a device beyond serving ads, and thus any permissions used by an ad library should be assumed to facilitate better ad delivery. Better ad delivery may encompass better user profiling or the efficient display of ads. Some permissions, such as \texttt{READ\_EXTERNAL\_STORAGE} and \texttt{WRITE\_EXTERNAL\_STORAGE}, may be useful for the caching of ads, which improves user experience. Other permission usage (such as that observed in~\cite{Grace:2012:UEA:2185448.2185464}), including reading a user's calendar, contact lists and call logs suggests highly invasive data collection. Moreover, the guile of some companies has been demonstrated by their use of the \texttt{RECORD\_AUDIO} permission to enable cross-device tracking using inaudible ultrasound~\cite{mavroudis2017privacy}.

Some libraries employ code obfuscation to frustrate reverse-engineering. This may be to make it difficult to commit ad fraud or disable ad functionality, but may also be used to hide unsavoury data collection practices. Moreover, many libraries also employ dynamic code loading whereby executable code is retrieved from the Internet and loaded by the library dynamically, thus defeating scalable static analysis~\cite{seo2016flexdroid}. The existence of dynamic programming techniques and obfuscation allow library developers to execute questionable code on devices while reducing the likelihood of being discovered. Already, potentially harmful code has been observed in libraries, affecting hundreds of thousands of apps across the smartphone ecosystem~\cite{seo2016flexdroid, 7546512}.

\subsection{Threat Model}

Our main adversaries are the ad networks that provide ad libraries, since they stand to benefit the most from invasive data collection. Invasive data collection is beneficial to both the advertiser and app developer (in the form of revenue), but the end-user suffers from an erosion in their privacy, and the fact that one or more third-parties are able to construct substantial profiles of their habits and interests. Other third-parties distributing libraries (that do not deliver ads) also stand to benefit if they can capture user data, since this data can be later sold. These other third-parties include malicious actors that attempt to introduce harmful code in libraries through repackaging, as observed in~\cite{7546512}.

We assume that our adversaries try to collect personal data, whether overtly or covertly, in order to facilitate their business objective. For ad networks, this business objective would be better user profiling for the purposes of targeting ads. For other non-advertising adversaries, this would be for collecting user data to sell or trade. Thus, in examining ILC, it is important to not focus solely on ad libraries. Additionally, it is important to understand that all streams of personal data may be valuable, as even innocuous pieces of data when combined may lead to greater privacy erosion through inferencing.

\subsubsection{What Personal Data is of Interest}

There are several streams of data on a smartphone that are of interest to adversaries. In what follows, we highlight what we suspect are the most important ones.\\

\noindent{\textbf{Location.} Most modern smartphones contain GPS hardware, which can precisely pinpoint a user's location. A coarse estimate of a user's location can also be obtained from other sources such as nearby Wi-Fi networks and cell towers. Invasive libraries can track user movements to determine where they live, work and socialise based on their location at various times of the week. Given movement patterns in the evenings, an adversary could infer that a user is an alcoholic. This information may be valuable to the user's insurance company, which itself may be inferred if the user has the insurance company app installed. Worryingly, the user's insurance app can even do this profiling itself.}\\

\noindent{\textbf{App Usage.} The list of apps installed on a device may be useful for understanding the interests of a user. Moreover, information about app usage can reveal the level of importance of each of these interests to the user. If app usage data is combined with location data, a broader picture may be painted of a user. For example, a user currently running a stopwatch app does not say much by itself, but when combined with coarse location data, an advertiser can determine a victim is likely training in the gym, as opposed to attending a social event, at a sports facility. This user may then be profiled and targeted with advertisements for protein powder, or more maliciously, anabolic steroids.}\\

\noindent{\textbf{Device Information.} Smartphones reveal device information such as device type and model. An adversary can use this to determine the amount of disposable income that a user has. Devices also contain unique identifier information such as an IMEI or SIM card information. An adversary tracking a user across devices (using SIM card information and assuming the user keeps their phone number) can be more confident in the disposable income of a user, if they are seen to use only high-end devices over time. Additionally, the IMEI of a device is also useful for tracking a user across apps, even if their device is restored to factory state.}\\

\noindent{\textbf{Communication.} Various communication data can be used to profile a user. From call logs and messages, an adversary can determine a user's close friends. With sentiment analysis, they may also be able to determine which contact is the user's spouse. Analysing text messages can also help to uncover a user's interests. The volume, duration, and time of phone calls can paint a picture of whether a user is very social and/or outgoing. Furthermore, parsing a victim's address book can allow the adversary to target persons with similar interests to the victim. Contacts in a user's address book may also prove useful if the user is blackmailed. For example, an attacker may threaten to notify a spouse of the victim's use of a dating app.}\\

\noindent{\textbf{Storage.} The files stored on a device may reveal other interests of a user. A user with more documents than pictures may be targeted with productivity tools instead of a digital camera. Conversely, a user with many pictures whose device is almost out of space may be targeted with ads for a new smartphone or memory cards. Further still, recently taken pictures combined with a user's location may suggest that the user enjoys photographing nature.}\\

\noindent{\textbf{Microphone.} The guile of adversaries has been demonstrated especially with the abuse of access to the device microphone. It is no longer necessary for an adversary to eavesdrop on a victim's conversations (although they may still do). Indeed, apps have been known to track users across devices using ultrasound~\cite{mavroudis2017privacy}.}\\
\section{Methodology}
\label{section-methodology}

To measure the danger of ILC in the real-world, we took the following steps:\\

\begin{enumerate}
\item Obtain all apps in the Google Play Store with more than one million downloads~(Section~\ref{section-app-lists}).
\item Identify the permissions that libraries within these apps are able to use~(Section~\ref{section-library-permission-usage}).
\item Use app lists from real-world devices to understand the number of libraries used within apps on devices and the permissions that they have access to~(Section~\ref{section-realworld-app-lists}).
\end{enumerate}

\subsection{Obtaining Apps for Analysis}
\label{section-app-lists}

In this study, we only considered apps available in the official Android app marketplace: the Google Play Store. Performing our analysis on this universe of apps, i.e., the entire Google Play Store, would require substantial storage and processing resources. For this reason, we instead opted to analyse popular apps. An app was considered to be popular if it had more than one million downloads. By analysing popular apps, we capture a large cross-section of those apps that would likely be installed on the average smartphone. This allows us to get a good indication of the threats to a majority of users, while limiting the amount of computing resources required for analysis. While additional information is lost by not considering less popular apps, these apps have a much smaller install-base and thus their contribution to ILC, if present, is already more limited.

To this end, we used a database of Google Play Store app metadata provided by the authors of~\cite{Taylor:2017:UUI:3052973.3052990} to identify and download all apps with more than one million downloads. This was~15,052 apps in total.

\subsection{Library Permission Usage}
\label{section-library-permission-usage}

Merely looking at the permissions recommended by library SDK documentation does not provide a complete understanding of permission usage within libraries. Indeed, libraries may use more permissions internally, and the app itself also needs to declare particular permissions (and have them granted) in order for the library to be able to use them. Conversely, just because an app declares (and has been granted) particular permissions does not mean that the library contains code that is able to leverage these permissions. Thus, several steps need to be taken to understand what permissions the libraries within each app have access to. In what follows, we outline the steps that were taken in this study to understand the permissions that libraries are able to use.\\

\begin{enumerate}
\item Decompile the \texttt{apk} file for an app to \texttt{smali} code using \texttt{apktool}~\cite{apktool}.
\item Use a whitelist of library signatures provided by the authors of FlexDroid~\cite{seo2016flexdroid} to identify library code as distinct from app code. Libraries could also be detected using techniques described in~\cite{Backes:2016:RTL:2976749.2978333}.
\item Use API-to-permission mappings from PScout~\cite{Au:2012:PAA:2382196.2382222} (improvement over Stowaway~\cite{Felt:2011:APD:2046707.2046779}) to understand what permissions are required by each API call observed in the \texttt{smali} code. At this point, we know what set of permissions can be used by library code and what set can be used by app code.
\item Take the intersection of the permissions declared in the app's manifest and the permissions observed in its \texttt{smali} code to determine what permissions libraries are actually able to leverage.
\end{enumerate}

Note that this technique may fail to reveal all permissions used if programming features such as dynamic code loading or reflection are employed. Thus, the permissions that are observed are a lower bound on the actual permissions that may be used in each case. For this reason, the results we obtain can also be considered as a lower-bound of the actual privacy risks coming from apps.

Not all apps were successfully decompiled. Indeed, a subset of them failed for reasons relating to technical shortcomings of the decompiler that was used. In the end, we were able to obtain library permission usage information for~14,976 apps.

\subsection{Obtaining Real-World App Lists}
\label{section-realworld-app-lists}

Central to the overall goal of understanding ILC is obtaining lists of installed apps on devices, hereafter called \textit{app lists}. To obtain app lists, we leveraged the Device Analyzer dataset. Device Analyzer is a project concerned with collecting usage statistics on smartphones. These usage statistics are collected and uploaded in the background by the client app, which is voluntarily installed by contributors to the project. Device Analyzer has data from over 30,000 contributors.

The most important data that we leverage from Device Analyzer is the app lists from each of the contributing devices and usage information for each of the apps in the app lists. Usage information tells when and how often apps are run by users. Concretely, we leveraged app lists and app usage information for 30,444 devices.

\subsection{Assumptions}

During our study we made several assumptions regarding the effect of run-time permissions and the effect of old Device Analyzer data.

\subsubsection{Assumption for Run-time Permissions}
\label{section-runtime-permissions}

The advent of run-time permissions has allowed users to selectively accept (or reject) permissions individually. For our study, we assume that apps are granted all permissions listed in their manifest as a matter of practicality, since the Device Analyzer dataset that we leverage currently does not capture run-time permissions that have been granted to apps. This may in turn cause us to overstate the number of permissions that have been granted to apps, and consequently, their libraries.

\begin{figure}[!t]
\centering
\includegraphics[width=3.25in]{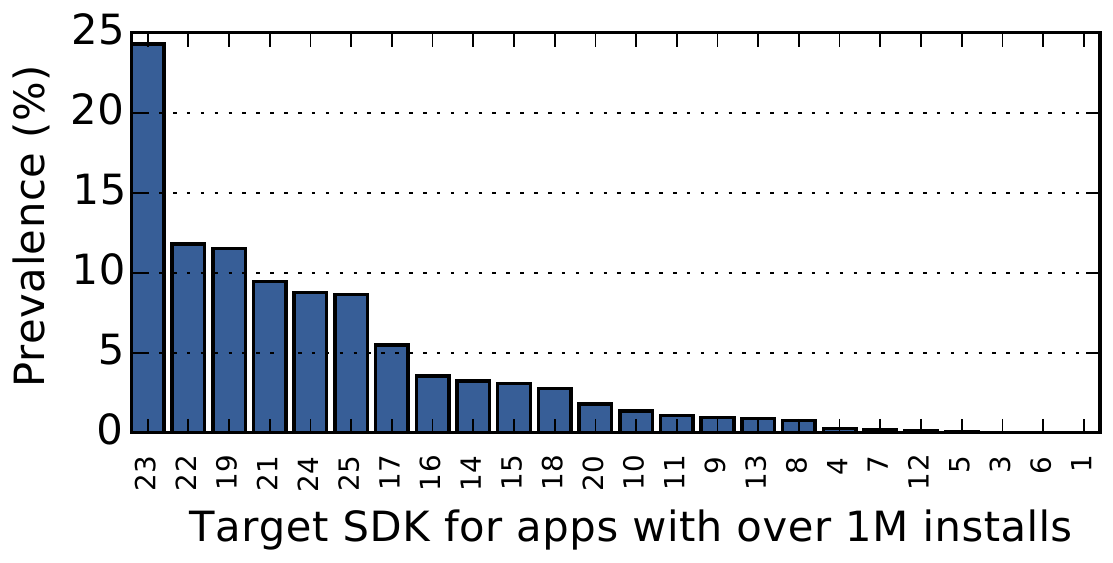}
\caption{API levels targeted by apps with more than one million downloads. Disproportionately more apps target API level~23, presumably to facilitate run-time permissions.}
\label{fig-targetsdk}
\end{figure}

However, for run-time permissions to be triggered, a device needs to be running at least Android~6.0 \textit{and} the app needs to have a \texttt{targetSDK} of~23 or higher. Fig.~\ref{fig-targetsdk} shows the \texttt{targetSDK} for the apps in our dataset. Approximately 60\% of apps have a \texttt{targetSDK} of~22 or lower, meaning that our assumption is valid, by default, in a majority of cases. 

Approximately~40\% of apps targeted API level~23 or higher. This means that 40\% of the apps in our dataset will trigger run-time permission requests provided that they are run on devices with Android 6.0 or higher. In reality, many devices are running older versions than Android~6.0, and thus run-time permissions will be triggered in less cases, though this number will increase as older devices get replaced with newer ones. Additionally, given user propensity to grant permissions~\cite{7427642, Felt:2012:APU:2335356.2335360}, we suspect that in other cases many run-time permissions will be granted. This leads to our assumption being valid in even more cases.

Disproportionately more apps target API level~23 as shown in Fig.~\ref{fig-targetsdk}. One explanation for this is that these app developers wanted to take advantage of run-time permissions and thus targeted this API level to implement the feature. It remains unclear, however, whether the developers of the approximately~60\% of apps targeting API level~22 or lower have incentive to switch to run-time permissions.

At this point, we remind the reader that these numbers describe the API levels targeted by popular apps. Popular apps have developers with financial incentives and resources to update their apps to target the newest API levels. Thus, we speculate that more than 60\% of unpopular apps will target API levels that pre-date run-time permissions, i.e., API level~22 or lower.

We used the Google Play Store metadata described in Section~\ref{section-app-lists} to make informed speculation on whether apps not supporting run-time permissions were likely to do so in the future. To infer this, we looked at the update history of apps not supporting run-time permissions. Approximately 64\% of apps not supporting run-time permissions have been updated since Android~6.0 was released (officially on October~5,~2015) and still fail to support run-time permissions. These apps, on average, received their latest update 353~days after the release of Android~6.0. The 36\% of apps that have not been updated since the release of Android~6.0 have an average update date of 378~days before the release of Android 6.0. This leads us to believe that many app developers simply have limited incentive to support run-time permissions at all. Thus we believe our assumption regarding run-time permissions is reasonable.

\subsubsection{Assumption for Old Device Analyzer Data}

Some data in Device Analyzer is old, and the apps used on devices in those datasets will be older versions than the versions we have downloaded and statically analysed. The apps could have since been updated to use different permissions and libraries. In performing our analysis, we treat old devices as if they were running the versions of the apps that we analysed. Given that apps are set to auto-update by default in the Google Play Store, we consider it a reasonable assumption that if the devices still had these apps installed they would be running the latest versions of said apps. Of course this is not always the case, but in the worst case what our assumption does is give an idea of the scale of potential ILC on devices that have the latest versions of those apps installed.

\section{Intra-Library Collusion (ILC)}
\label{section-ilc}

To make measurements of ILC meaningful, it is first important to quantify the number of distinct libraries detected within apps on the devices in our dataset. Within our dataset, each device had an average of~23.6 detectable libraries within the popular apps that were installed on those devices.

\begin{table}[!t]
\centering
\caption{Most popular libraries detected within apps with more than one million downloads. Note that we omit libraries detected in less than 1\% of apps.}
\label{table-most-popular-libraries}
\begin{tabular}{p{1.75in}x{1.3in}}
\hline
Library & \% of apps \\ \hline
com/facebook & 11.9 \\ \hline
com/google/android/gms/analytics & 9.8 \\ \hline
com/flurry & 6.3 \\ \hline
com/chartboost/sdk & 5.9 \\ \hline
com/unity3d & 5.2 \\ \hline
com/applovin & 3.5 \\ \hline
com/mopub & 3.1 \\ \hline
com/inmobi & 3.0 \\ \hline
com/google/ads & 3.0 \\ \hline
com/google/android/gcm & 2.7 \\ \hline
com/tapjoy & 2.4 \\ \hline
org/cocos2d & 2.4 \\ \hline
com/amazon & 2.0 \\ \hline
com/millennialmedia & 1.6 \\ \hline
org/apache/commons & 1.4 \\ \hline
com/heyzap & 1.4 \\ \hline
com/nostra13/universalimageloader & 1.3 \\ \hline
com/adobe/air & 1.0 \\ \hline
\end{tabular}
\end{table}

The Top~18 most popular libraries (i.e., libraries observed in more than~1\% of apps that were studied) within apps in our dataset is shown in Table~\ref{table-most-popular-libraries}. The most popular libraries included Facebook, Google Analytics, Flurry, and Chartboost. Popular ad libraries such as InMobi, MoPub, Millennial Media, Heyzap and TapJoy were also seen in the Top~18 most popular libraries. Several utility libraries were also popular, providing functionality such as loading/caching images (\texttt{com/nostra13/universalimageloader}) and rendering graphics (\texttt{com/unity3d} and \texttt{org/cocos2d}).

\subsection{Which Libraries may Benefit from ILC}

\begin{figure}[!t]
\centering
\includegraphics[width=3.25in]{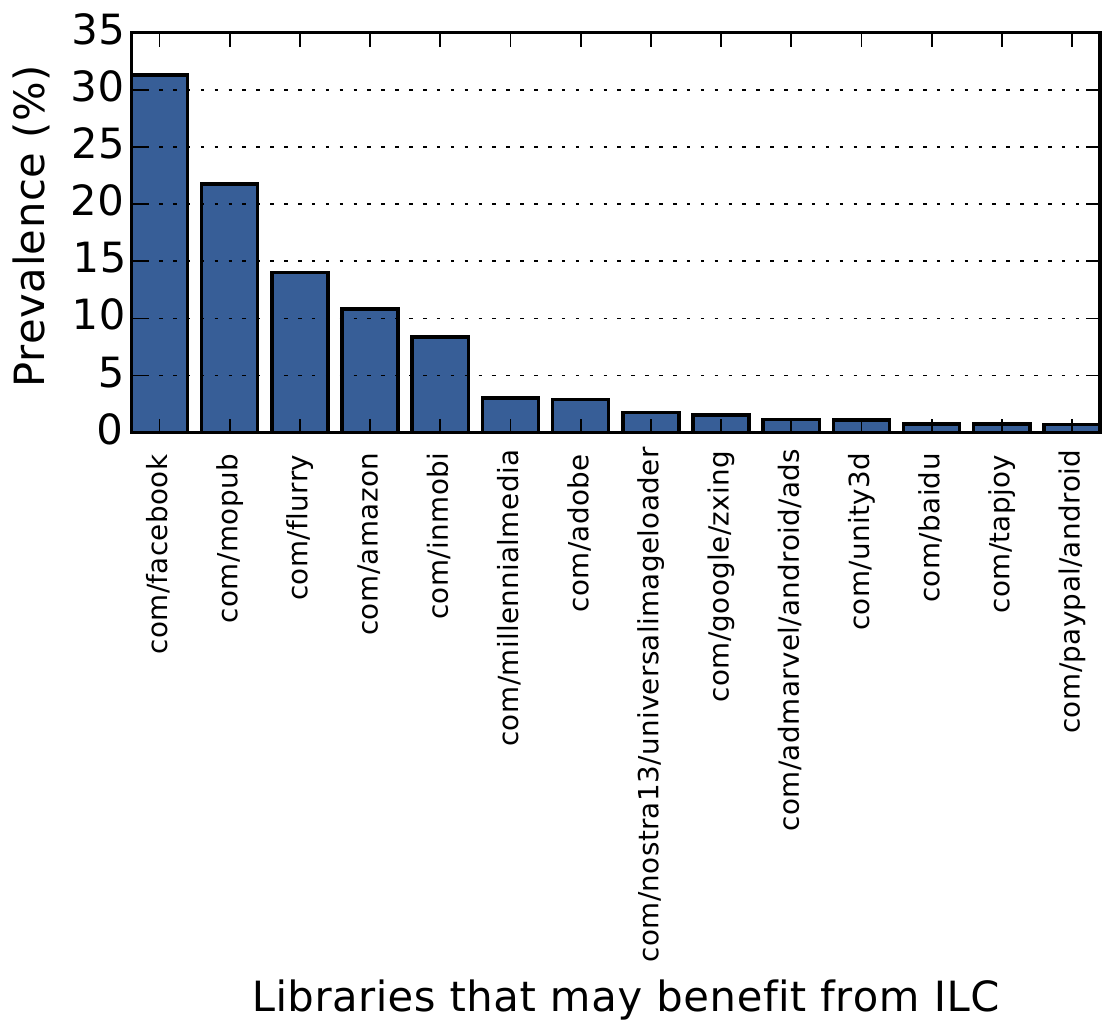}
\caption{Libraries that are potentially able to benefit from ILC. For clarity, libraries appearing less than 0.5\% of the time are omitted.}
\label{fig-which-libraries-empowered}
\end{figure}

Fig.~\ref{fig-which-libraries-empowered} shows which libraries are potentially able to benefit from ILC. This was calculated as the number of each library potentially able to benefit from ILC divided by the total number of instances where libraries were potentially able to benefit from ILC. In most cases, it was \texttt{com/facebook} that was potentially able to benefit from ILC with 31.3\%. Other libraries with the potential to benefit from ILC were \texttt{com/mopub} (21.8\%), \texttt{com/flurry} (14.0\%), \texttt{com/amazon} (10.8\%), and \texttt{com/inmobi} (8.4\%).

Worryingly, the Top~5 libraries that are potentially able to benefit from ILC include MoPub, Flurry Analytics and InMobi; known advertising/analytics providers. Note that this observation is not an indictment against any of the libraries mentioned. Rather, it shows the extent to which ILC could be leveraged in the real-world by these libraries if they had the desire to do so. Given the fierce competition between ad libraries, it is conceivable that wily ad networks would exploit ILC by aggregating user data on their servers, maximising profits while evading detection.

\subsection{How Libraries may Benefit from ILC}

To fully understand the potential risk to users, we measured the number of distinct libraries per device that had the potential to benefit from ILC. These results are given in Table~\ref{table-num-libraries-ilc}. Approximately two in five devices (42.4\%) are not susceptible to an ILC attack. We note, however, that a device can go from not being susceptible to being susceptible through the installation of a single app. On the other hand, 57.6\% of devices had one or more libraries that are potentially able to benefit from ILC. In fact, one in five (20.4\%) devices in our dataset had three or more libraries that would be able to benefit from ILC. This is equivalent to approximately~6,000 devices in our dataset, but translated to the real-world this would amount to hundreds of millions of devices. At this scale, even a slight improvement in ad targeting gained by leveraging ILC would reap substantial revenue for ad networks having the requisite guile.

\begin{table}[]
\centering
\caption{Number of libraries per device that had increased access to permissions.}
\label{table-num-libraries-ilc}
\begin{tabular}{x{1.55in}x{1.5in}}
\hline
Number of Libraries & \% of devices \\ \hline
0  & 42.4 \\ \hline
1  & 20.7 \\ \hline
2  & 16.5 \\ \hline
3  & 10.5 \\ \hline
4  & 5.8  \\ \hline
5+ & 4.1  \\ \hline
\end{tabular}
\end{table}

In addition to understanding the number of libraries able to benefit from ILC, it is also important to quantify what benefit they would receive by doing so. To this end, we examined the number of additional permissions that a library leveraging ILC would be able to obtain. This result is given in Table~\ref{table-num-permissions-ilc}. To measure the increase in permissions, the difference between the total number of permissions obtained using ILC and the maximum number of permissions the library could access from any one app on the device was taken. In most cases (69.2\%), libraries would be able to access one additional permission if they leveraged ILC. The number of devices monotonically decreased as the number of permissions increased. Worryingly, libraries would obtain access to three or more permission on 9.6\% of devices if they leveraged ILC. While 9.6\% is not a significant number of devices, we remind the reader that this would translate to hundreds of millions of devices in the real-world.

\begin{table}[]
\centering
\caption{Number of additional permissions a library had access to beyond the single-app maximum.}
\label{table-num-permissions-ilc}
\begin{tabular}{x{1.55in}x{1.5in}}
\hline
Number of Permissions & \% of cases \\ \hline
1  & 69.2 \\ \hline
2  & 21.2 \\ \hline
3  & 5.9 \\ \hline
4  & 2.8 \\ \hline
5+ & 0.9 \\ \hline
\end{tabular}
\end{table}
\section{How the Potential for ILC has Evolved in Two Years}
\label{section-longitudinal}

Taylor and Martinovic~\cite{Taylor:2017:UUI:3052973.3052990} conducted the first systematic large-scale measurement of how permission usage in apps has increased over time. They also quantified the extent to which embedded libraries benefited from this permission increase in a phenomenon they call \textit{library empowerment}. Along similar lines, we examine how the potential for ILC has changed over time as a result of libraries and apps using more permissions as they are updated.

To perform this measurement, we used a freely available dataset~\cite{archiveplaydrone} of historical versions of apps compiled using the PlayDrone tool~\cite{Viennot:2014:MSG:2591971.2592003}. The historical versions of apps used were the versions as they were available in the Google Play Store in October~2014. By measuring the potential for ILC in these versions of apps as we did in Section~\ref{section-ilc}, we can see how the potential for ILC has changed over an approximately two-and-a-half year period.

Not all~14,976 apps in our current dataset were available in the historic dataset of apps. This is because some apps currently in our dataset did not exist in October~2014. Overall, we were able to obtain~11,836 historic apps. Due to technical shortcomings of the decompilers used, we were able to obtain library permission usage information for~11,821 apps. The difference in number of apps in old and new datasets may skew the results of our longitudinal study. To prevent such skew, we took the intersection of the sets of apps that were available in the old and new datasets. In the end, we were able to do a longitudinal study of how the potential for ILC attacks on devices changed for~11,774 of apps (with more than one million downloads) over a two-and-a-half year period.

Note that increases in the potential for ILC are not only caused by the apps themselves using more permissions. Indeed, the libraries embedded within apps may have now been updated to introduce additional code that makes use of permission-protected APIs. Additionally, wider adoption of particular libraries by app developers may cause substantial changes in the potential for ILC, especially if these libraries use many permissions.

\subsection{Libraries that May Exploit ILC}

\begin{figure}[!t]
\centering
\includegraphics[width=3.25in]{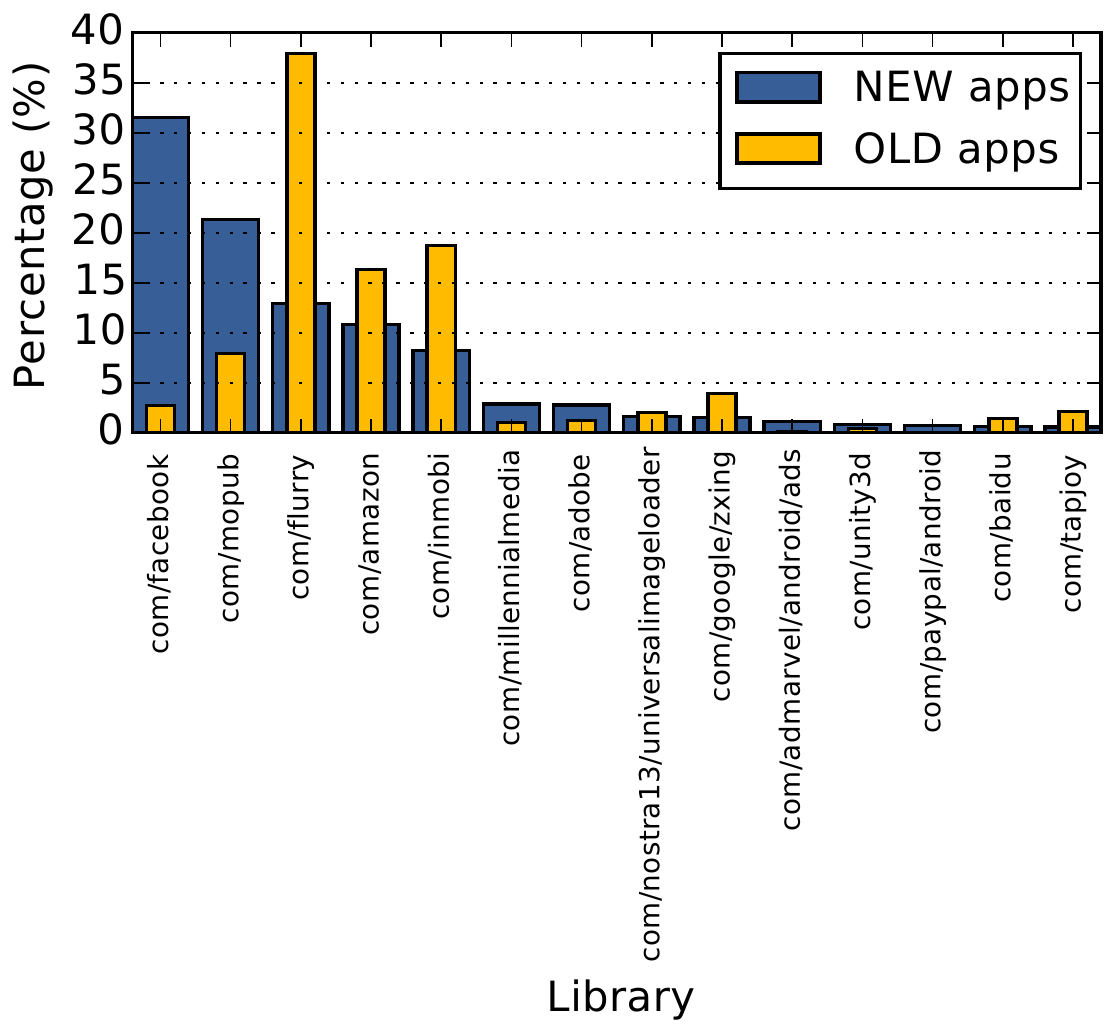}
\caption{Longitudinal look at changes in the libraries that are able to benefit from ILC. For clarity, libraries appearing less than 0.5\% of the time are omitted.}
\label{fig-longitudinal-ilc}
\end{figure}

Fig.~\ref{fig-longitudinal-ilc} looks at the longitudinal changes in the libraries that are able to exploit ILC. We use \texttt{NEW} to refer to results generated from analysing the new versions of apps, and \texttt{OLD} refers to results obtained by analysing historical versions of apps. The Facebook library had the largest increase, going from 2.7\% to 31.5\%. The Flurry library had the largest decrease going from 37.9\% to 12.9\%.

We manually investigated the four libraries that had the most significant changes over the period to understand why this was the case. These libraries were \texttt{com/facebook}, \texttt{com/mopub}, \texttt{com/flurry} and \texttt{com/inmobi}. The increase in prevalence for \texttt{com/facebook} came about from increased numbers of apps using the library, as well as the fact that the permissions used by the library increased. For \texttt{com/mopub}, the reason was the same: more apps started using the library and its permission usage also increased. On the side of decreases over time, \texttt{com/flurry} and \texttt{com/inmobi} had their decrease because of less apps using the library. For both libraries, no change in number of permissions used was detected. These changes demonstrate the extent to which the possibility of ILC across devices can change simply because libraries start using one or more new permissions and/or because they become more popular.

\subsection{Changes in Potential Benefit from ILC}

Table~\ref{table-num-libraries-ilc-longitudinal} summarises our longitudinal analysis of the number of libraries per device that were able to exploit ILC. Worryingly, there was a 19.7\% decrease in the number of libraries per device that were unable to benefit from ILC. This suggests that library and permission usage evolution over time is facilitating increases in the potential to exploit ILC. There was also a 13.8\% decrease in the case where one library was able to benefit from ILC. The number of cases where two or more libraries per device were able to benefit from ILC went from 21.5\% of cases to 35.5\% of cases. Thus, not only is the potential for ILC increasing, but the consequences of the attack in terms of number of libraries that can benefit are increasing as well.

\begin{table}[]
\centering
\caption{Longitudinal look at the number of libraries per device that had increased access to permissions.}
\label{table-num-libraries-ilc-longitudinal}
\begin{tabular}{x{1.1in}x{0.55in}x{0.55in}r}
\hline
Number of Libraries & OLD \% & NEW \% & \% Change \\ \hline
0  & 53.8 & 43.2 & -19.7\%  \\ \hline
1  & 24.7 & 21.3 & -13.8\%  \\ \hline
2  & 12.5 & 16.4 & +31.2\%  \\ \hline
3  &  5.6 & 10.1 & +80.4\%  \\ \hline
4  &  2.1 &  5.3 & +152.4\% \\ \hline
5+ &  1.3 &  3.7 & +184.6\% \\ \hline
\end{tabular}
\end{table}

We further analysed the number of additional permissions that libraries would be able to exploit if they leveraged ILC. The results of this analysis is shown in Table~\ref{table-num-permissions-ilc-longitudinal}. The cases where libraries were able to leverage one additional permission fell from 86.5\% to 68.5\%. Worryingly, the cases where libraries could access two or more new permissions, increased from 13.5\% of cases to 31.5\% of cases, a 133\% increase. Once again, it was the more dangerous case (increase of~2+ permissions) that had the greater increase with the less dangerous case (increase of 1~permission) having a decrease.

\begin{table}[]
\centering
\caption{Longitudinal look at the number of additional permissions a library had access to beyond the single-app maximum.}
\label{table-num-permissions-ilc-longitudinal}
\begin{tabular}{x{1.2in}x{0.5in}x{0.5in}r}
\hline
Number of Permissions & OLD \% & NEW \% & \% Change \\ \hline
1  & 86.5\% & 68.5\% & -20.8\% \\ \hline
2+ & 13.5\% & 31.5\% & +133.3\% \\ \hline
\end{tabular}
\end{table}
\section{How Often is Sensitive Data Sent to Ad Library Servers}
\label{section-frequency-leaks}

We now turn our attention to a case study on ad libraries. This is because ad libraries have financial incentive to exploit ILC and from our measurements, several ad libraries were among the top libraries that were able to benefit from ILC. Manual decompilation and analysis of the binary \texttt{JAR} files of several ad libraries revealed that sensitive data such as location data and nearby Wi-Fi networks are routinely sent with each ad request.

For this reason, we wanted to approximate a lower bound on how often and to how many different ad networks such sensitive personal data is sent. This is useful for understanding the quantity of data that is exfiltrated from a device to ad networks per day. With this estimate, one can obtain a more comprehensive understanding of the extent to which ad libraries already collect information suitable for exploiting ILC.\\

To make the calculations, we needed to understand how many (and how often) apps were run on devices per day. This information was obtained by analysing data on app usage in the Device Analyzer dataset. In what follows, we discuss some practical assumptions that were made in performing our calculations.\\

\noindent{\textbf{Leakage of personal data.} We assume that each ad library sends personal data (if relevant permissions are available) with each ad request. From the sample of ad libraries that we manually analysed, this assumption is correct. The literature~\cite{stevens2012investigating, Grace:2012:UEA:2185448.2185464} also agrees that ad libraries typically send data useful for profiling with each ad request. Throughout our calculations, however, we were careful to deem an ad library as sending sensitive data with ad requests only if:\\}

\begin{enumerate}
\item The ad library contains permission-protected Android API calls.
\item The app embedding the library has declared the relevant permissions that allow the ad library to make these API calls.
\end{enumerate}

\noindent{\textbf{Frequency of leakage.} In reality, apps are used multiple times a day and many of these apps contain ad libraries which fetch and continuously update ads. Unfortunately, Device Analyzer data is not granular enough to say exactly how many times per day and app is launched or which activity within an app is run. This fine-grained data (were it available) would enable us to precisely determine how often ads were shown. This is because it is possible to determine whether an ad is embedded within a particular activity of an app. Given the absence of the requisite fine-grained data, we assume that a single ad is loaded, i.e. sensitive data is sent once, per ad library per app per day \textit{if} the app is run by the user. Anecdotal data suggests that this is a gross underestimation of what actually happens. However, in giving this baseline, we obtain an estimate of the lower bound of sensitive data leakage by ad libraries. Future researchers with better estimates of how many ads are served per ad library per app per day will be able to scale our measurements accordingly.}\\

In making our calculations, we only considered ad libraries that had a prevalence of more than 1\% across apps in our dataset. Across the Device Analyzer dataset, devices used an average of 7.4~third-party apps (available in the Google Play Store) per day. Of these apps,~4.7 of them were in our app dataset (i.e., they had one million downloads or more). From these~4.7 apps, a mean of~2.4 sensitive data leakages were caused by ad libraries per device per day, using our assumptions. That is, approximately 50\% of apps with over one million downloads leak sensitive data per device per day on average. Further analysis of the data reveals that sensitive data was leaked to approximately~1.7 different ad servers per device per day.

At the upper end of the spectrum, one device had~132 instances of sensitive data leakages per day. Interestingly, at least one device sent data to all the ad servers that were considered in our measurement. We discuss these observations further and put them into context in Section~\ref{section-discussion-frequency}.
\section{Discussion}

We have measured the potential for ILC to take place and the benefits to be gained by libraries exploiting ILC. We have also shown how the potential for ILC has increased over a two-and-a-half year period. This increase is facilitated by apps using more permissions but also the libraries themselves being updated to use more permission-protected API calls internally.

Prior work, and our own observations, confirm that libraries already send enough data back to their servers to facilitate an ILC attack. This does not mean, however, that any of the libraries we identified actually engage in this practice. Given the nature of ILC and the fact that data aggregation happening on the server-side is opaque, it becomes very difficult to know whether ILC happens in practice. This problem is further complicated by the fact that libraries may have legitimate reasons to explain why particular pieces of personal data need to be transmitted off a device. Once off the device, however, any aggregation that happens would be invisible.

We suspect that the advertising and analytics industry has the most to gain from exploiting ILC. Given the guile shown by ad libraries and ad networks in general, we believe that this may be a very attractive attack, especially considering that it would be hard to prove that it was happening. Given the fierce competition in the advertising and analytics space, any additional signals about users that can be leveraged from data that is already being collected can improve an ad network's targeting potential. Even if this improvement is a small one, when translated to the app ecosystem of millions of apps and billions of devices, ILC has the potential to generate (or is already generating) a windfall for ad networks.

The main catalyst that allows ILC to happen is the failure of the Android permission system to separate the privileges of libraries and their host apps. However, even if this privilege separation were to be implemented in future Android versions (using strategies such as those highlighted in Section~\ref{section-privilege-separation}), it may be the case as with run-time permissions (as shown in Section~\ref{section-runtime-permissions}) that there is limited incentive for app developers to support it. On the contrary, there may be incentive for app developers to \textit{not} support library privilege separation, as it may impact their profits negatively.

\subsection{Addressing the Problem of ILC}

In this section, we discuss technical and non-technical approaches that can be used to mitigate the problem of ILC.

\subsubsection{Technical Approaches}

Prior work, such as those highlighted in Section~\ref{section-privilege-separation}, attempt to separate privileges and thus alleviate the problem of ILC somewhat. However, we now use the example of ad libraries to demonstrate that several concerns will remain if these strategies are used.

If privileges are removed from libraries altogether, then ad libraries will have more difficulty in targeting ads to users. This increases the likelihood that advertisers and ad networks will not be interested in such systems. Additionally, app developers stand to lose revenue as well and thus may be uninterested in implementing privacy-preserving features. Moreover, depending on the privilege separation approach used, user data may be explicitly passed from apps to libraries using data passing APIs first observed by Book and Wallach~\cite{Book:2013:CCS:2516760.2516762} (see Section~\ref{section-data-collection}) and confirmed by us.

While good for user privacy, less privileged ad libraries may actually harm the ecosystem. Much of the effort that goes into developing apps comes from the expectation of the app developers that they will receive a return on their investment when they monetise their apps. Poor app monetisation could serve as a deterrent to new app developers entering the market and thus the end users may ultimately suffer from reduced content.

\subsubsection{Non-technical Approaches}

Major app stores, such as Google Play, could attempt to limit ILC by means of modifying their developer policies. Nation states may also enact and enforce laws that prohibit the aggressive cross-aggregation of data that happens when ILC is exploited. These steps go in the right direction, but violations may be difficult to detect and enforce in practice.

The first problem is detecting that a library employs ILC. The only positive evidence that might be available is that libraries leak personal data off a device. While this is a necessary condition for ILC to happen, it is not sufficient. Moreover, detecting privacy leaks in the first place is challenging. Static and dynamic analysis tools are not perfect and may fail to detect all leaks in apps.

Even with perfect detection of leaks, libraries may have legitimate reasons for sending data off a device, and so merely observing leakage of sensitive user data does not imply guilt in exploiting ILC. Thus the major challenge is born out of the fact that the actual data aggregation during ILC happens on third-party servers. That is, the actual point of maliciousness happens where it is not transparent from the outside. Thus, app stores and regulatory bodies have an uphill battle. One promising avenue to infer the exploitation of ILC by third-parties is the use of differential analysis techniques, such as those employed by L{\'e}cuyer et al.~\cite{Lecuyer:2014:XEW:2671225.2671229}

The second problem is that of enforcement. While apps may be in violation of the terms and conditions of app stores (or indeed even local and national law), there is no unifying framework for enforcement across the app ecosystem. Indeed, there are a number of third-party app stores that provide apps to millions of users, which may not necessarily impose the required privacy policies. Moreover, apps may need to be penalised on a case by case basis, since they may only be in violation of a law if the app is downloaded by a user in a particular country, and there exists proof that specific types of data on the user have been aggregated. The aforementioned challenges make the large-scale mitigation of ILC difficult in practice.

\subsection{Frequency of Data Leakage}
\label{section-discussion-frequency}

The frequency of data leakage by libraries in apps is a source of added concern. In Section~\ref{section-frequency-leaks}, it was observed that the average device studied had~2.4 leakages of sensitive data by ad libraries per day. This quantity of leakages comes from the~7.4 third-party apps used in a day, of which~4.7 have more than one million downloads. Generalising, approximately 50\% of popular apps installed on devices leak sensitive data from these devices to ad networks.

We would like to remind the reader at this point that our assumptions with regard to the frequency of data leakage were very conservative. Indeed, we assumed one sensitive data leak per ad library per app per day, only if:\\

\begin{enumerate}
\item An app containing an ad library was seen to be running on a device in a day.
\item An ad library within the running app has the ability to use one or more permissions \textit{and} the permissions declared by the app allowed it to do so.
\end{enumerate}

Many apps are run by users more than once per day. Moreover, ad libraries also rotate the ads that are shown while an app is being run, thus adding to the number of ad requests leaking data. Thus our estimates fail to account for increased data leakage along two dimensions.

Unfortunately, the data provided by Device Analyzer, while rich in many regards, does not allow us to take any more accurate measurements. Moreover, we only considered apps with more than one million downloads. This caused us to ignore approximately 36\% of ``unpopular'' third-party apps that were run on devices per day, further reducing our estimate.

For these reasons, we consider our estimates of data leakage per device per day to be very conservative. However, even if these conservative numbers represent what happens in the real-world, it is not reassuring that ad libraries have the capability to send all the sensitive data that they have access to from a device more than twice a day and to almost two different ad networks. We leave more accurate measurements of the frequency of data leakages by ad libraries on real-world devices as an interesting area for future work.

\subsection{Limitations of this Study}

In this section, the limitations of our methodology are highlighted and discussed.

\subsubsection{Library Permission Usage}

Our static analysis approach to determine library permission usage, as detailed in Section~\ref{section-library-permission-usage}, is limited in its ability to handle dead code and dynamic code. This is an inherent limitation of static analysis approaches. If our system identifies permission usage in dead code, it will incorrectly attribute it to the library in question and overstate the library's permission usage. On the other hand, since our system fails to handle dynamic code, it may miss permission usage and thus understate the library's permission usage.

Our anecdotal observations, however, suggest that dead code and dynamic code are not a significant fraction of the code contained within libraries. Moreover, each type of code has the opposite effect on our estimation of library permission usage. For this reason, we believe that our estimates generally give a good representation of the scale of the problem. We leave handling dead code and dynamic code as an interesting area for future work.

\subsubsection{Device Analyzer Data}

The Device Analyzer client app is more likely to be installed by a technical set of users, since it is an academic endeavour and has mostly been promoted indirectly through publications that make use of its data. For this reason, app lists and app usage information in Device Analyzer may not fully reflect that of the average user in the real-world.

However, we suspect that since contributors to Device Analyzer are more technical, our results actually underestimate the results that would be seen if a more representative sample of users was used. This is because we imagine technical users are more savvy and more likely to take additional steps to preserve their privacy when choosing apps and using their devices.
\section{Related Work}

Related work in the area can be divided into several categories: permission usage increases in apps and libraries, data collection by libraries, approaches to mitigating privacy leaks, and sensitive data tracking. In what follows, we discuss the most closely related work in each category.

\subsection{Permission Usage Increases}

Taylor and Martinovic~\cite{Taylor:2017:UUI:3052973.3052990} first measured the extent to which apps were adding permissions over time across the Google Play Store. Before that, there was a wide belief that systematic permission increases were happening but without concrete data. The authors found that apps on a whole had a tendency to add permissions over time, and importantly, that embedded libraries were also benefiting from these permissions. They refer to this phenomenon as ``library empowerment''. Along similar lines, we look at the totality of personal data that libraries would be able to aggregate by leveraging ILC. Additionally, we use a similar longitudinal study to understand how the potential for libraries exploiting ILC has changed over time.

Book et al.~\cite{DBLP:journals/corr/abs-1303-0857} explore the extent to which Android ad libraries expanded their permission usage over time. By looking at app release dates and the ad library versions that the apps were using, they were able to build a chronological map of ad library permission usage. They found increases in permission usage over time as well as observed that several of the new permissions added posed risks to user privacy.

When taken together with the aforementioned work of Taylor and Martinovic, it paints a worrying picture in that apps are using more permissions, while at the same time libraries are becoming more able to exploit these additional permissions to their benefit. Complementary to this, we measure the extent of what invasive libraries would be able to gain, were they to exploit their combined privileges using ILC.

\subsection{Data Collection by Libraries}
\label{section-data-collection}

A number of authors looked at the collection and transmission of personal data by libraries. Grace et al.~\cite{Grace:2012:UEA:2185448.2185464} developed a system called AdRisk, and used it to examine potential privacy risks posed by ad libraries. They found that most of the ad libraries studied collected private information such as a user's location. Worryingly, they found evidence of invasive data collection taking place such as accessing a user's call logs. The authors also found that some ad libraries downloaded and executed code dynamically, leading to security risks on devices.

Stevens et al.~\cite{stevens2012investigating} also looked at user privacy in ad libraries. The authors discovered unsavoury practices being performed by ad libraries such as the probing of permissions to see which ones were available to it. The permissions checked for were beyond the required and optional permissions specified by the library documentation. This demonstrates the guile of ad libraries in accessing and trying to access user data. Presumably, the ad libraries would then access data guarded by permissions that they have access to. The authors confirm their suspicion by using network traces from a major network provider to confirm the existence of private data leaked by the ad libraries in question. Given these observations that ad libraries will go to extreme (and overt) lengths for better user profiling, our work aims to understand what libraries may be doing covertly.

Along similar lines, Book and Wallach~\cite{Book:2013:CCS:2516760.2516762} consider the situation where apps themselves pass user data to ad libraries through internal ad library APIs\footnote{In our manual analysis, we also observed such API calls in ad libraries. Additionally, ad library SDKs also provided documentation on how app developers should use these APIs. We can thus confirm the observations of these authors.}. The authors did a study on 114,000 apps and found that app popularity is correlated with privacy leakage. The authors argue that the marginal increase in revenue gained, when taken across millions of users, seems to incentivise the violation of user privacy. With this motivation from ad networks in mind, we aim to measure the potentially more damaging threat of ILC. We also add our voice to the call for greater privilege separation between apps and their libraries.

\subsection{Privilege Separation}
\label{section-privilege-separation}

A number of authors propose privilege separation strategies for apps and their embedded libraries. Shekhar et al.~\cite{Shekhar:2012:ASS:2362793.2362821} propose AdSplit, an extension to Android that allows apps and ad libraries to run as separate processes with separate UIDs. AdSplit is able to automatically separate libraries from their host apps. It does this by decompiling an app and replacing ad library code with a ``stub library'' supporting the same API. This stub library can then call a separate AdDroid advertisement service. Similar to AdSplit, AFrame achieves process and permission isolation, but also provides display and input isolation~\cite{Zhang:2013:AIA:2523649.2523652}.

Pearce et al.~\cite{Pearce:2012:APS:2414456.2414498} take a different approach to solving the problem. They build support for advertising directly into the Android platform with a service called AdDroid. This eliminates the need for ad networks to provide ad libraries at all. Instead, AdDroid relays the transmission of data and handles displaying ads, obviating the need for an ad library to be resident in app code.

A major drawback of the privilege separation systems described above is the willingness of ad networks to adopt them. To counter this, Liu and Simpson~\cite{liu2016privacy} propose privacy-preserving targeted mobile advertising (PPTMA). Their goal is a system that provides privacy guarantees to users, while at the same time being palatable to ad networks. PPTMA acts as middleware that is positioned between untrusted ad libraries and sensitive user data. PPTMA hooks privacy-sensitive APIs and takes different actions depending on whether an ad network is ``cooperative'' or whether an app is whitelisted, and the like.

Seo et al.~\cite{seo2016flexdroid} took a more general approach and focused on privilege separation between apps and their libraries as a whole, i.e., their focus was not solely ad libraries. They devise a system called FlexDroid, which provides dynamic and fine-grained privilege separation for third-party libraries. FlexDroid is an extension to the Android permission system that allows app developers to specify different sets of permissions for different libraries. It, however, remains unclear whether app developers would be interested in limiting permissions to ad libraries, since it would directly affect their profit margins.

Similar to FlexDroid, Compac provides fine-grained access control at the component level~\cite{Wang:2014:CEC:2557547.2557560}. NativeGuard provides similar protection against libraries written in native code~\cite{Sun:2014:NPA:2627393.2627396}. Finally, Roesner et al.~\cite{180364} explore the secure embedding of interfaces, a technique that would be useful to support ad libraries.

\subsection{Flow and Taint Tracking}

Enck et al.~\cite{Enck:2010:TIT:1924943.1924971} took a very early look at how apps used and transmitted private data using a system they developed called TaintDroid. Their system is an extension to the Android platform that allows the tracking of sensitive data flows through apps. Using TaintDroid, the authors found 68 instances of data leaks across~20 of the~30 apps that they studied. In a half of the studied apps, the authors found instances where location data was sent to several advertisement servers, without user consent, and even in cases where no ad was displayed.

Arzt et al.~\cite{Arzt:2014:FPC:2594291.2594299} propose a system called FlowDroid that does static taint analysis of Android apps. By accurately modelling the Android life cycle, the authors were able to obtain high precision in their tests. Along similar lines, Wei et al.~\cite{Wei:2014:APG:2660267.2660357} propose Amandroid, a static analysis tool for vetting Android apps. Both FlowDroid and Amandroid can be used to statically determine whether data leakage can happen in apps. These tools can assist in quickly identifying apps that may have concerning behaviour. Apps seen to be sending sensitive data off devices may then be further analysed and their libraries subjected to differential analysis~\cite{Lecuyer:2014:XEW:2671225.2671229} to attempt to identify the exploitation of ILC.
\section{Conclusion}

In this paper, we highlighted a novel and dangerous privilege escalation vulnerability called intra-library collusion. ILC is enabled by inherent shortcomings in the Android permission model whereby privileges between apps and their embedded libraries are not separated. If exploited, ILC allows libraries to secretly aggregate multiple sources of sensitive user data by leveraging the permissions that they have been granted within two or more apps. Using app lists from over~30,000 devices, we observed that several popular social, advertising, and analytics libraries are able to exploit ILC. Some~57.6\% of the devices studied had at least one library that could benefit from ILC. These libraries could access a two or more additional permissions in~30.8\% of cases.

By doing a longitudinal study, we observed that several libraries increased their ability to exploit ILC. This happened because the libraries themselves started to use additional permissions or they became more popular. The capabilities gained by libraries exploiting ILC were also seen to increase over time. By doing a case study on ad libraries, we showed that ad libraries leak sensitive data from a device up to~2.4 times a day and that the average user has their personal data sent to~1.7 different ad servers per day.

As apps and smart devices become more popular, it is important to ensure that user privacy is protected from those who attempt to profit from it. By highlighting the novel capability of adversaries in the form of ILC, we highlight the potential danger, and add our voice to the call for privilege separation between apps and their libraries. This work takes us a step further in securing user privacy as smartphones and smart devices become ubiquitous and more ingrained in our lives.



\bibliographystyle{abbrv}
\bibliography{deviceanalyzer}

\begin{thebibliography}{10}

\bibitem{apktool}
{Apktool - A tool for reverse engineering 3rd party, closed, binary Android
  apps}.
\newblock https://ibotpeaches.github.io/Apktool/.

\bibitem{androiddangerousperms}
{Requesting Permissions - Android Developers}.
\newblock https://developer.android.com/guide/topics/permissions/
  requesting.html.

\bibitem{archiveplaydrone}
archive.org.
\newblock {Android Apps}.
\newblock https://archive.org/details/android\_apps.

\bibitem{Arzt:2014:FPC:2594291.2594299}
S.~Arzt, S.~Rasthofer, C.~Fritz, E.~Bodden, A.~Bartel, J.~Klein, Y.~Le~Traon,
  D.~Octeau, and P.~McDaniel.
\newblock {FlowDroid: Precise Context, Flow, Field, Object-sensitive and
  Lifecycle-aware Taint Analysis for Android Apps}.
\newblock In {\em Proceedings of the 35th ACM SIGPLAN Conference on Programming
  Language Design and Implementation}, PLDI '14, pages 259--269, New York, NY,
  USA, 2014. ACM.

\bibitem{Au:2012:PAA:2382196.2382222}
K.~W.~Y. Au, Y.~F. Zhou, Z.~Huang, and D.~Lie.
\newblock {PScout: Analyzing the Android Permission Specification}.
\newblock In {\em Proceedings of the 2012 ACM Conference on Computer and
  Communications Security}, CCS '12, pages 217--228, New York, NY, USA, 2012.
  ACM.

\bibitem{Backes:2016:RTL:2976749.2978333}
M.~Backes, S.~Bugiel, and E.~Derr.
\newblock {Reliable Third-Party Library Detection in Android and Its Security
  Applications}.
\newblock In {\em Proceedings of the 2016 ACM SIGSAC Conference on Computer and
  Communications Security}, CCS '16, pages 356--367, New York, NY, USA, 2016.
  ACM.

\bibitem{DBLP:journals/corr/abs-1303-0857}
T.~Book, A.~Pridgen, and D.~S. Wallach.
\newblock {Longitudinal Analysis of Android Ad Library Permissions}.
\newblock {\em CoRR}, abs/1303.0857, 2013.

\bibitem{Book:2013:CCS:2516760.2516762}
T.~Book and D.~S. Wallach.
\newblock {A Case of Collusion: A Study of the Interface Between Ad Libraries
  and Their Apps}.
\newblock In {\em Proceedings of the Third ACM Workshop on Security and Privacy
  in Smartphones and Mobile Devices}, SPSM '13, pages 79--86, New York, NY,
  USA, 2013. ACM.

\bibitem{Bosu:2017:CDL:3052973.3053004}
A.~Bosu, F.~Liu, D.~D. Yao, and G.~Wang.
\newblock {Collusive Data Leak and More: Large-scale Threat Analysis of
  Inter-app Communications}.
\newblock In {\em Proceedings of the 2017 ACM on Asia Conference on Computer
  and Communications Security}, ASIA CCS '17, pages 71--85, New York, NY, USA,
  2017. ACM.

\bibitem{bugiel2012towards}
S.~Bugiel, L.~Davi, A.~Dmitrienko, T.~Fischer, A.-R. Sadeghi, and B.~Shastry.
\newblock {Towards Taming Privilege-Escalation Attacks on Android}.
\newblock In {\em NDSS}, 2012.

\bibitem{Chan:2012:DAA:2185448.2185466}
P.~P. Chan, L.~C. Hui, and S.~M. Yiu.
\newblock {DroidChecker: Analyzing Android Applications for Capability Leak}.
\newblock In {\em Proceedings of the 5th ACM Conference on Security and Privacy
  in Wireless and Mobile Networks}, WISEC '12, pages 125--136, New York, NY,
  USA, 2012. ACM.

\bibitem{7546512}
K.~Chen, X.~Wang, Y.~Chen, P.~Wang, Y.~Lee, X.~Wang, B.~Ma, A.~Wang, Y.~Zhang,
  and W.~Zou.
\newblock {Following Devil's Footprints: Cross-Platform Analysis of Potentially
  Harmful Libraries on Android and iOS}.
\newblock In {\em IEEE Symposium on Security and Privacy}, IEEE S\&P, pages
  357--376, May 2016.

\bibitem{7427642}
N.~Eling, S.~Rasthofer, M.~Kolhagen, E.~Bodden, and P.~Buxmann.
\newblock {Investigating Users' Reaction to Fine-Grained Data Requests: A
  Market Experiment}.
\newblock In {\em 2016 49th Hawaii International Conference on System Sciences
  (HICSS)}, pages 3666--3675, Jan 2016.

\bibitem{Enck:2010:TIT:1924943.1924971}
W.~Enck, P.~Gilbert, B.-G. Chun, L.~P. Cox, J.~Jung, P.~McDaniel, and A.~N.
  Sheth.
\newblock {TaintDroid: An Information-flow Tracking System for Realtime Privacy
  Monitoring on Smartphones}.
\newblock In {\em Proceedings of the 9th USENIX Conference on Operating Systems
  Design and Implementation}, OSDI'10, pages 393--407, Berkeley, CA, USA, 2010.
  USENIX Association.

\bibitem{Fang:2016:RCA:2897845.2897914}
Z.~Fang, W.~Han, D.~Li, Z.~Guo, D.~Guo, X.~S. Wang, Z.~Qian, and H.~Chen.
\newblock {revDroid: Code Analysis of the Side Effects After Dynamic Permission
  Revocation of Android Apps}.
\newblock In {\em Proceedings of the 11th ACM on Asia Conference on Computer
  and Communications Security}, ASIA CCS '16, pages 747--758, New York, NY,
  USA, 2016. ACM.

\bibitem{Felt:2011:APD:2046707.2046779}
A.~P. Felt, E.~Chin, S.~Hanna, D.~Song, and D.~Wagner.
\newblock {Android Permissions Demystified}.
\newblock In {\em Proceedings of the 18th ACM Conference on Computer and
  Communications Security}, CCS '11, pages 627--638, New York, NY, USA, 2011.
  ACM.

\bibitem{Felt:2012:APU:2335356.2335360}
A.~P. Felt, E.~Ha, S.~Egelman, A.~Haney, E.~Chin, and D.~Wagner.
\newblock {Android Permissions: User Attention, Comprehension, and Behavior}.
\newblock In {\em Proceedings of the 8th Symposium on Usable Privacy and
  Security (SOUPS 2012)}, SOUPS '12, pages 3:1--3:14, New York, NY, USA, 2012.
  ACM.

\bibitem{Felt:2011:PRA:2028067.2028089}
A.~P. Felt, H.~J. Wang, A.~Moshchuk, S.~Hanna, and E.~Chin.
\newblock {Permission Re-delegation: Attacks and Defenses}.
\newblock In {\em Proceedings of the 20th USENIX Conference on Security},
  SEC'11, pages 22--22, Berkeley, CA, USA, 2011. USENIX Association.

\bibitem{ndsssdcl}
M.~Grace, Y.~Zhou, Z.~Wang, and X.~Jiang.
\newblock {Systematic Detection of Capability Leaks in Stock Android
  Smartphones}.
\newblock In {\em Network and Distributed System Security Symposium (NDSS
  '12)}, 2012.

\bibitem{Grace:2012:UEA:2185448.2185464}
M.~C. Grace, W.~Zhou, X.~Jiang, and A.-R. Sadeghi.
\newblock {Unsafe Exposure Analysis of Mobile In-app Advertisements}.
\newblock In {\em Proceedings of the 5th ACM Conference on Security and Privacy
  in Wireless and Mobile Networks}, WISEC '12, pages 101--112, New York, NY,
  USA, 2012. ACM.

\bibitem{Lecuyer:2014:XEW:2671225.2671229}
M.~L{\'e}cuyer, G.~Ducoffe, F.~Lan, A.~Papancea, T.~Petsios, R.~Spahn,
  A.~Chaintreau, and R.~Geambasu.
\newblock {XRay: Enhancing the Web's Transparency with Differential
  Correlation}.
\newblock In {\em Proceedings of the 23rd USENIX Conference on Security
  Symposium}, SEC'14, pages 49--64, Berkeley, CA, USA, 2014. USENIX
  Association.

\bibitem{liu2016privacy}
Y.~Liu and A.~Simpson.
\newblock {Privacy-preserving targeted mobile advertising: requirements, design
  and a prototype implementation}.
\newblock {\em Software: Practice and Experience}, 46(12):1657--1684, 2016.

\bibitem{Lu:2012:CSV:2382196.2382223}
L.~Lu, Z.~Li, Z.~Wu, W.~Lee, and G.~Jiang.
\newblock {CHEX: Statically Vetting Android Apps for Component Hijacking
  Vulnerabilities}.
\newblock In {\em Proceedings of the 2012 ACM Conference on Computer and
  Communications Security}, CCS '12, pages 229--240, New York, NY, USA, 2012.
  ACM.

\bibitem{mavroudis2017privacy}
V.~Mavroudis, S.~Hao, Y.~Fratantonio, F.~Maggi, C.~Kruegel, and G.~Vigna.
\newblock {On the Privacy and Security of the Ultrasound Ecosystem}.
\newblock {\em Proceedings on Privacy Enhancing Technologies}, 2017(2):95--112,
  2017.

\bibitem{nielsonmanyapps}
Nielson.
\newblock {So Many Apps, So Much More Time for Entertainment}.
\newblock
  http://www.nielsen.com/us/en/insights/news/2015/so-many-apps-so-much-more-time-for-entertainment.html,
  June 2015.

\bibitem{Pearce:2012:APS:2414456.2414498}
P.~Pearce, A.~P. Felt, G.~Nunez, and D.~Wagner.
\newblock {AdDroid: Privilege Separation for Applications and Advertisers in
  Android}.
\newblock In {\em Proceedings of the 7th ACM Symposium on Information, Computer
  and Communications Security}, ASIACCS '12, pages 71--72, New York, NY, USA,
  2012. ACM.

\bibitem{180364}
F.~Roesner and T.~Kohno.
\newblock {Securing Embedded User Interfaces: Android and Beyond}.
\newblock In {\em Presented as part of the 22nd USENIX Security Symposium
  (USENIX Security 13)}, pages 97--112, Washington, D.C., 2013. USENIX.

\bibitem{Seneviratne:2014:PUT:2636242.2636244}
S.~Seneviratne, A.~Seneviratne, P.~Mohapatra, and A.~Mahanti.
\newblock {Predicting User Traits from a Snapshot of Apps Installed on a
  Smartphone}.
\newblock {\em SIGMOBILE Mobile Computing and Communications Review},
  18(2):1--8, June 2014.

\bibitem{seo2016flexdroid}
J.~Seo, D.~Kim, D.~Cho, T.~Kim, and I.~Shin.
\newblock {FlexDroid: Enforcing In-App Privilege Separation in Android}.
\newblock In {\em NDSS}, 2016.

\bibitem{Shekhar:2012:ASS:2362793.2362821}
S.~Shekhar, M.~Dietz, and D.~S. Wallach.
\newblock {AdSplit: Separating Smartphone Advertising from Applications}.
\newblock In {\em Proceedings of the 21st USENIX Conference on Security
  Symposium}, Security'12, pages 28--28, Berkeley, CA, USA, 2012. USENIX
  Association.

\bibitem{spensky2016sok}
C.~Spensky, J.~Stewart, A.~Yerukhimovich, R.~Shay, A.~Trachtenberg, R.~Housley,
  and R.~K. Cunningham.
\newblock {SoK: Privacy on Mobile Devices - It's Complicated}.
\newblock {\em Proceedings on Privacy Enhancing Technologies}, 2016(3):96--116,
  2016.

\bibitem{stevens2012investigating}
R.~Stevens, C.~Gibler, J.~Crussell, J.~Erickson, and H.~Chen.
\newblock {Investigating User Privacy in Android Ad Libraries}.
\newblock In {\em Workshop on Mobile Security Technologies (MoST)}, page~10,
  2012.

\bibitem{Sun:2014:NPA:2627393.2627396}
M.~Sun and G.~Tan.
\newblock {NativeGuard: Protecting Android Applications from Third-party Native
  Libraries}.
\newblock In {\em Proceedings of the 2014 ACM Conference on Security and
  Privacy in Wireless \& Mobile Networks}, WiSec '14, pages 165--176, New York,
  NY, USA, 2014. ACM.

\bibitem{Taylor:2016:SSP:2994459.2994474}
V.~F. Taylor and I.~Martinovic.
\newblock {SecuRank: Starving Permission-Hungry Apps Using Contextual
  Permission Analysis}.
\newblock In {\em Proceedings of the 6th ACM CCS Workshop on Security and
  Privacy in Smartphones and Mobile Devices (SPSM 2016)}, SPSM '16, pages
  43--52, New York, NY, USA, 2016. ACM.

\bibitem{Taylor:2017:UUI:3052973.3052990}
V.~F. Taylor and I.~Martinovic.
\newblock {To Update or Not to Update: Insights From a Two-Year Study of
  Android App Evolution}.
\newblock In {\em Proceedings of the 2017 ACM on Asia Conference on Computer
  and Communications Security}, ASIA CCS '17, pages 45--57, New York, NY, USA,
  2017. ACM.

\bibitem{Ur:2012:SUS:2335356.2335362}
B.~Ur, P.~G. Leon, L.~F. Cranor, R.~Shay, and Y.~Wang.
\newblock {Smart, Useful, Scary, Creepy: Perceptions of Online Behavioral
  Advertising}.
\newblock In {\em Proceedings of the Eighth Symposium on Usable Privacy and
  Security}, SOUPS '12, pages 4:1--4:15, New York, NY, USA, 2012. ACM.

\bibitem{Viennot:2014:MSG:2591971.2592003}
N.~Viennot, E.~Garcia, and J.~Nieh.
\newblock {A Measurement Study of Google Play}.
\newblock In {\em The 2014 ACM International Conference on Measurement and
  Modeling of Computer Systems}, SIGMETRICS '14, pages 221--233, New York, NY,
  USA, 2014. ACM.

\bibitem{Wagner:2015:DAP:2766498.2774992}
D.~T. Wagner, D.~R. Thomas, A.~R. Beresford, and A.~Rice.
\newblock {Device Analyzer: A Privacy-aware Platform to Support Research on the
  Android Ecosystem}.
\newblock In {\em Proceedings of the 8th ACM Conference on Security \& Privacy
  in Wireless and Mobile Networks}, WiSec '15, pages 34:1--34:2, New York, NY,
  USA, 2015. ACM.

\bibitem{Wang:2014:CEC:2557547.2557560}
Y.~Wang, S.~Hariharan, C.~Zhao, J.~Liu, and W.~Du.
\newblock {Compac: Enforce Component-level Access Control in Android}.
\newblock In {\em Proceedings of the 4th ACM Conference on Data and Application
  Security and Privacy}, CODASPY '14, pages 25--36, New York, NY, USA, 2014.
  ACM.

\bibitem{Wei:2014:APG:2660267.2660357}
F.~Wei, S.~Roy, X.~Ou, and Robby.
\newblock {Amandroid: A Precise and General Inter-component Data Flow Analysis
  Framework for Security Vetting of Android Apps}.
\newblock In {\em Proceedings of the 2014 ACM SIGSAC Conference on Computer and
  Communications Security}, CCS '14, pages 1329--1341, New York, NY, USA, 2014.
  ACM.

\bibitem{Zhang:2013:AIA:2523649.2523652}
X.~Zhang, A.~Ahlawat, and W.~Du.
\newblock {AFrame: Isolating Advertisements from Mobile Applications in
  Android}.
\newblock In {\em Proceedings of the 29th Annual Computer Security Applications
  Conference}, ACSAC '13, pages 9--18, New York, NY, USA, 2013. ACM.

\end{thebibliography}

\end{document}